\newcount\Comments  
\documentclass[fleqn,usenatbib]{mnras}

\usepackage{newtxtext,newtxmath}
 
\usepackage[T1]{fontenc}
\usepackage{float}
\usepackage[justification=centering]{caption}
\DeclareRobustCommand{\VAN}[3]{#2}
\let\VANthebibliography\thebibliography
\def\thebibliography{\DeclareRobustCommand{\VAN}[3]{##3}\VANthebibliography}

\usepackage{graphicx}	
\usepackage{amsmath}	
\usepackage{amssymb}	
\usepackage{gensymb}
\usepackage[
singlelinecheck=false 
]{caption}
\usepackage{threeparttable}
\usepackage{float}
\usepackage{color}
\definecolor{darkgreen}{rgb}{0,0.5,0}
\definecolor{purple}{rgb}{1,0,1}
\newcommand{\kibitz}[2]{\ifnum\Comments=1\textcolor{#1}{#2}\fi}

\newcommand{\per}{$^{-1}$}

\title[MeerKAT's view of the Bullet cluster]{MeerKAT's View of the Bullet Cluster 1E 0657-55.8}

\author[S.P. Sikhosana et al.]{
S.P. Sikhosana,$^{1,2}$\thanks{E-mail: sikhosanas@ukzn.ac.za (S.P.S.)}
K. Knowles,$^{3,4}$
M. Hilton,$^{1,2,6}$
K. Moodley,$^{1,2}$ 
and M. Murgia,$^{5}$
\\
$^{1}$Astrophysics Research Centre, University of KwaZulu-Natal, Durban, 3696, South Africa\\
$^{2}$ School of Mathematics, Statistics $\&$ Computer Science, University of KwaZulu-Natal, Westville Campus,
Durban 4041, South Africa\\
$^{3}$Department of Physics and Electronics, Rhodes University, P.O. Box 94, Makhanda 6140, South Africa\\
$^{4}$South African Radio Astronomy Observatory, 2 Fir Street, Observatory, Cape Town 7405, South Africa\\
$^{5}$INAF-Osservatorio Astronomico di Cagliari, Via della Scienza 5, 09047 Selargius, IT\\
$^{6}$Wits Centre for Astrophysics, School of Physics, University of the Witwatersrand, Private Bag 3, 2050, Johannesburg, South Africa\\
}

\date{Accepted XXX. Received YYY; in original form ZZZ}

\pubyear{2022}

\begin{document}
\label{firstpage}
\pagerange{\pageref{firstpage}--\pageref{lastpage}}
\maketitle

\begin{abstract}
The Bullet cluster (1E\,0657-55.8) is a massive merging system at redshift $z$ = 0.296, known to host a powerful radio halo and a relic. Here we present high fidelity MeerKAT $L$-band (0.9 $-$ 1.7\,GHz) observations of the Bullet cluster, where we trace a larger extent of both the halo and relic in comparison to previous studies. The size of the recovered halo is 1.6\,Mpc $\times$ 1.3\,Mpc and the largest linear size of the relic is $\sim$988\,kpc. We detect a new decrement feature on the southern outskirts of the halo emission, where a region appears to have a lower surface brightness in comparison to its surroundings. The larger extension on the outskirts of the halo is faint, which suggests lower relativistic electron density or a weaker magnetic field. An in-band spectral index map of the halo reveals radial steepening towards the edges, likely due to synchrotron electron ageing. The integrated spectral index of the radio halo is 1.1\;$\pm$\;0.2. We perform a radio$-$X-ray surface brightness point-to-point analysis, which reveals a linear correlation for the radio halo. This indicates that the halo emission is produced by primary re-acceleration mechanisms. Finally, we derive a radio Mach number of $\mathcal{M}_R$= 4.6\;$\pm$\;0.9 for the relic shock region, which is higher than the Mach number inferred by earlier analyses based on X-ray data. Discrepancies between radio and X-ray Mach numbers have been observed for multiple systems, studies suggest that this is due to various factors, including relic orientation.
\end{abstract}




\section{Introduction}
Radio observations of galaxy clusters reveal cluster-scale diffuse synchrotron emission that is not associated with individual galaxies, but rather with the intracluster medium (ICM). The formation of these radio sources is linked to major merger activity in the ICM \citep{2014IJMPD..2330007B}. Hence, studying the properties of these diffuse radio sources gives insight into astrophysical processes such as cosmic ray transportation and particle (re)acceleration in the ICM. The extended diffuse radio sources can be categorised into radio halos (RHs) and radio relics based on their size, morphology and location \citep[for reviews]{2012A&ARv..20...54F,2019SSRv..215...16V}.  
\par
Giant radio halos (GRHs) which are $\sim$Mpc in size, located in central regions of clusters, and with low polarisation percentages, have two main formation theories. The first is the secondary `hadronic' model, in which electrons originate from hadronic collisions between the long-living relativistic protons in the ICM and thermal ions \citep[e.g.,][]{1980BAAS...12..471D,2008MNRAS.385.1211P,2010arXiv1011.0729K,2011A&A...527A..99E}. This formation theory has not been widely accepted due to the lack of observational evidence of gamma rays in clusters, which are a by-product of the hadronic processes \citep{2003ApJ...588..155R,2010PhRvD..82i2004A,2018ApJ...860...85A}. The second model is the primary `re-acceleration' model. According to this model, a pool of pre-existing electrons \citep{2017MNRAS.465.4800P} is re-accelerated through second-order Fermi mechanism by ICM turbulence developing during cluster mergers \citep{2001MNRAS.320..365B,2005MNRAS.357.1313C,2012A&ARv..20...54F,2014MNRAS.443.3564D}. Observational studies have linked GRHs to host clusters with high mass (M$_{500}$ > 4$\times 10^{14}$M$_{\odot}$) and X-ray and/or optical merger signatures \citep{2005ApJ...627..733M,2013ApJ...777..141C,2014MNRAS.440.2901S,2014ApJ...786...49L,2015A&A...579A..92K,2019MNRAS.486.1332K,2021MNRAS.500.2236R}. Studies of large populations of radio halos show that these sources are not ubiquitous \citep{1982A&A...116..137H,2001ApJ...548..639K}. Their radio power is correlated with their X-ray properties \citep{2000astro.ph.12166L,2015ApJ...813...77Y}, the largest linear size \citep[LLS;][]{2009A&A...507.1257G}, and the integrated inverse Compton parameter \citep{2012MNRAS.421L.112B}. Studies of samples derived from Sunyaev-Zel'dovich \citep[SZ;][]{1972CoASP...4..173S} surveys  have shown that there is a higher occurrence rate of radio halos compared to the rate predicted by previous studies of X-ray selected samples \citep{2012MNRAS.421L.112B,2014MNRAS.437.2163S,2015A&A...580A..97C,2022A&A...660A..78B}. Although the re-acceleration theory is accepted as the primary mechanism, there are still a few aspects that need further investigation. The main open question is the source of the pre-existing population of electrons. The suggested sources include electrons deposited in the ICM by active galactic nuclei (AGN) activity, thermal protons pre-accelerated by cluster merger shocks, and the hadronic secondary electrons \citep{2005MNRAS.363.1173B,2016MNRAS.459..277S,2017SciA....3E1634D}. Another mystery is the presence of radio halos in clusters that show no trace of merger activity \citep{2017A&A...603A.125V,2019MNRAS.486L..80K}. Other models, such as magnetic re-connection, attempt to resolve such constraints by proposing hybrid models \citep{2016MNRAS.458.2584B}. Various observational phenomena can be used to distinguish formation theories, one such example is the radio-X-ray point-to-point correlation study \citet{2001A&A...369..441G,2019A&A...628A..83C,2020ApJ...897...93B,2021A&A...646A.135R}. 
\par
Relics are arc-shaped, $\sim$Mpc scale, highly polarised sources ($\gtrsim$ 20$\%$) that are located at clusters' peripheral regions. Observations by \citet{1979ApJ...233..453J, 2012MNRAS.426...40B,2016MNRAS.460L..84B} and \citet{2020MNRAS.tmpL..75L} show that their origin is linked to shock waves generated in the ICM by merger events \citep{1998A&A...332..395E,1999ApJ...518..603R}. However, the underlying particle acceleration mechanism is still under debate \citep{2019MNRAS.489.3905S}. In the mechanism of first-order diffusive shock acceleration \citep[DSA;][]{1978MNRAS.182..443B,1983RPPh...46..973D,1991SSRv...58..259J}, cosmic-ray protons and electrons are assumed to be accelerated from the thermal pool up to relativistic energies at the cluster merger shocks. Although this mechanism can explain the general properties of relic emission, several observational features remain unexplained \citep{2019MNRAS.489.3905S}, such as the non-detection of gamma rays in clusters that host relics \citep{2014MNRAS.437.2291V}, and the low Mach numbers observed in shocks \citep{2014IJMPD..2330007B,2020A&A...634A..64B}. The second mechanism proposes the re-acceleration of fossil relativistic electrons via DSA at the cluster shocks \citep{2005ApJ...627..733M,2007MNRAS.375...77H,2013MNRAS.435.1061P,2017ApJ...840...42K}. This mechanism reproduces the observed spectrum \citep{2014MNRAS.444.3130D,2015MNRAS.449.1486S,2017NatAs...1E...5V} and does not require shocks to have large Mach numbers, as the pre-existing electrons have enough energy to be re-accelerated to relativistic speeds \citep{2014IJMPD..2330007B}. However, this mechanism also presents challenges as there are expected phenomena that are yet to be observed \citep{2019MNRAS.489.3905S}. For example, the connection between active galactic nuclei (AGN; candidate seed electron source) and radio relics could be established only in few cases \citep{2014ApJ...785....1B,2017ApJ...835..197V}.
\par
The Bullet cluster (1E 0657-55.8) is a massive (11.4$^{+2.2}_{-1.8}\times$10$^{14}\;M_\odot$) galaxy cluster at redshift $z$=0.296 which known to host both a GRH and a relic \citep{2000ApJ...544..686L,2015MNRAS.449.1486S}. The cluster is reported to have recently experienced a major merger \citep{2002ApJ...567L..27M,2006ESASP.604..723M}. The observed collision is approximately perpendicular to the line of sight, which makes this cluster an intriguing case study for various wavelengths. At radio wavelengths, MeerKAT \citep{2009IEEEP..97.1522J} is an excellent instrument for observing this cluster because the densely populated core increases sensitivity to large scale diffuse emission. MeerKAT is a precursor of the Square Kilometre Array. It is one of the most sensitive radio arrays in the pre-SKA era \citep{2018ApJ...856..180C,2020ApJ...888...61M}. The full MeerKAT array consists of 64 antennas, with 70$\%$ of the dishes located at the core ($<$400 m) of the array. The minimum baseline for the array is 29 m and the maximum baseline is 7.7 km. The array configuration makes MeerKAT an excellent instrument for studying faint extended diffuse emission in galaxy clusters. Extracting the extended emission requires short baselines, while long baselines are needed to disentangle compact source emission. Both the requirements are simultaneously available with MeerKAT's full array observations. Additionally, MeerKAT's high sensitivity significantly reduces the integration time required to detect faint diffuse emission. As a result of superior sensitivity, particularly in the core, shorter observations from MeerKAT detect faint features in the diffuse emission that similar instruments require significantly longer integration times to reproduce. The Bullet cluster was observed by MeerKAT as part of its Galaxy Cluster Legacy Survey \citep[MGCLS;][]{2021arXiv211105673K}, which observed 115 clusters at $L$-band during the science verification and early science exploitation of MeerKAT. We independently reduced the data and present the results. 
\par
The paper is organised as follows. In Section \ref{sec:multiwave} we summarise previous multi-wavelength studies of the Bullet cluster. We outline our MeerKAT observations and data reduction in Section \ref{sec:observe}. In Section \ref{sec:results} we present the MeerKAT results, and in Section \ref{sec:discuss} we discuss the radio vs X-ray surface brightness correlation and radio Mach number of the relic. We adopt a $\Lambda$CDM flat cosmology with $H_0 = 70$\,kms{\per}Mpc{\per}, $\Omega_m = 0.3$, and $\Omega_\Lambda = 0.7$. At the redshift of the Bullet cluster ($z$ = 0.296), the luminosity distance is 1529\,Mpc and 1$\arcsec$ corresponds to 4.413\,kpc.

\section{MULTI-WAVELENGTH OBSERVATIONS OF 1E 0657-55.8}
\label{sec:multiwave}
\begin{table}
	\centering
	\caption{ Multiwavelength properties of 1E 0657$-$55.8. SZ values are from the ACT DR5 catalog \citep{2021ApJS..253....3H}. The X-ray properties are from \citet{2010ApJ...723.1523M}. }
	\label{tab:bullet}
	\begin{tabular}{lc} 
		\hline
		R.A.$_{\rm J2000}$ (hh:mm:ss.s) & 06:58:32.7  \\
		Dec.$_{\rm J2000}$ (dd:mm:ss.s) & $-$55:57:19.0 \\
		redshift & 0.296 \\
		$Y_{\rm 500,SZ} \; (10^{-4} \rm arcmin^2)$ & 13.3 \;$\pm$\; 0.4 \\
		$M_{\rm 500,SZ} \; (10^{14}\;M_\odot)$ & 11.4$^{+2.2}_{-1.8}$ \\
		$L_{\rm 500,X} \; (10^{44}$ erg s\per) & 20.69 \;$\pm$\; 0.05 \\
		$T_{\rm X} (keV)$ & 13.56 \;$\pm$\; 0.14 \\
		$Z_{\rm X} \; (Z_\odot)$ & 0.36 \;$\pm$\; 0.02 \\
		\hline
	\end{tabular}
\end{table}

\citet{1995ApJ...444..532T,1998ApJ...496L...5T} were the first to detect the Bullet cluster using the \textit{ROSAT} satellite \citep{1992eocm.rept....9V}, the Advanced Satellite for Cosmology and Astrophysics \citep{1994PASJ...46L..37T}, and the Einstein IPC instrument \citep{1979ApJ...230..540G}. \citet{1998ApJ...496L...5T} also measured that the Bullet cluster has a spectroscopic redshift of $z=0.296$. There have since been numerous multi-wavelength follow-ups of this cluster \citep{2000ApJ...544..686L,2002ApJ...567L..27M,2004ApJ...604..596C}. \citet{2002A&A...386..816B} used spectroscopic data from the  ESO New Technology Telescope  \citep{1989SPIE.1114..302T} to study the dynamical state of the cluster. Their results showed that the system has a sub-cluster which is offset from the main cluster position by  0.7\,Mpc in the western direction. They found that the sub-cluster had a virial mass of 1.3$\times$10$^{13}$M$_{\odot}$ and the main cluster had virial mass of 1.3$\times$10$^{15}$M$_{\odot}$. They concluded that the mass of the sub-cluster was significantly larger at the pre-merger stage. Hence, they concluded that the Bullet cluster has recently (within the past 0.15\,Gyr) experienced a major merger. \citet{2004ApJ...604..596C} used Very Large Telescope (VLT FORS1) data to reconstruct weak lensing maps of the Bullet cluster. Their weak lensing and X-ray comparison studies revealed that the X-ray centroid peak and reconstructed cluster mass were misaligned, hence showing that the total mass of the system does not trace the baryonic mass. This result indicated that dark matter is collisionless. Other weak lensing studies have further solidified these findings \citep{2006ApJ...648L.109C,2006ApJ...652..937B,2016A&A...594A.121P}.
\par
\textit{Chandra} X-ray studies by \citet{2002ApJ...567L..27M} and \citet{2006ESASP.604..723M} showed that the merger was a `textbook' example of a bow shock merger. They measured the Mach number of the bow shock to be $\mathcal{M}_X$ = 3.0 \;$\pm$\; 0.4. The velocity of the cluster was found to be 3000$-$4000\,km/s. Their observations also indicate that the sub-cluster traversed the main cluster 0.1$-$0.2\,Gyrs ago and is in its final stage of being destroyed by dynamic gas instabilities. They studied the temperature profile across the shock and found that the electrons are fast heated at the bow shock front. \citet{2019A&A...628A.100D} used Atacama Large Millimeter/submillimeter Array and Atacama Compact Array SZ observations to study the bow shock in the cluster. They found an SZ-derived Mach number of $\mathcal{M}_{SZ}$ = 2.53$^{+0.33}_{-0.25}$, in agreement with their X-ray-derived value of $\mathcal{M}_{X}$ = 2.57 \;$\pm$\; 0.23, obtained after reprocessing archival \textit{Chandra} data.  
\par
\citet{2000ApJ...544..686L} were the first to report the detection of a radio halo hosted by the Bullet cluster. They observed the cluster using the Australia Telescope Compact Array (ATCA) at  1.3, 2.4, 4.9, 5.9, and 8.8\,GHz. There have since been follow-up studies of the radio emission using deeper, and higher frequency, ATCA observations \citep{2014MNRAS.440.2901S,2015MNRAS.449.1486S,2016Ap&SS.361..255M}. The first detection of a luminous `toothbrush' relic was reported by \citet{2015MNRAS.449.1486S}. They measured an integrated flux density of 77.8 \;$\pm$\; 3.1 and 4.8 \;$\pm$\; 0.6 mJy, $\alpha$\footnote{S$_{\nu}\, \propto \, \nu^{\alpha}$} was $-$1.07 \;$\pm$\; 0.03 and $-$1.66 \;$\pm$\; 0.14 for regions A and B, respectively (see Figure \ref{fig:bulletreg}), and P$_{1.4GHz}$ = (2.3 \;$\pm$\; 0.1)$\times$10$^{25}$\,WHz$^{-1}$. Their X-ray analysis revealed that the relic coincides with a shock opposite the western bow shock. The Mach number of the shock was measured to be $\mathcal{M}_X$ = 2.5$^{+1.8}_{-0.8}$. The multi-wavelength properties of the Bullet cluster are summarised in Table \ref{tab:bullet}.

\section{MeerKAT Observations and Data Reduction}
\label{sec:observe}
The presence of a radio halo and relic in the Bullet cluster \citep{2000ApJ...544..686L,2014MNRAS.440.2901S,2015MNRAS.449.1486S,2016Ap&SS.361..255M} made it an excellent science verification target for MeerKAT. The Bullet cluster was observed on 24 June 2018 using 61 antennas. The data were recorded at $L$-band, which has a total bandwidth of 856\,MHz divided into 4096 channels, and the integration time was set to eight seconds. The full observation was $\sim$11 hours, including the overheads. For the observations, the primary flux calibrators were J0408$-$6545 and J1331$+$3030. J0408$-$6545 was observed for 10 minutes after every 1-hour scan of the target and gain calibrator, and J1331$+$3030  was observed for 10 minutes for the last hour scans. The secondary gain calibrators were J0825$-$5010 and PKS\,0647$-$475. The gain calibrators were each observed for two minutes after every 20-minute target scan. The observation details are summarised in Table \ref{tab:observations}. The raw data was provided to us by the Observatory in
2018 as part of a PhD project. This observation was later included in the MGCLS's first data release \citep{2021arXiv211105673K}. However, for this analysis, we reduced this data independently of the MGCLS. The MGCLS DR1 only provides image products, we needed to further process the data by removing the compact sources in the visibility plane and re-image at low resolution. Our independent processing also allows for sanity checking and comparison against the Obit processing used in MGCLS DR1. The details of the data reduction are discussed below.

\subsection{Data Reduction}
\label{subsec:reduce}
The data were reduced using \textsc{oxkat} v1.0\footnote{\url{https://github.com/IanHeywood/oxkat}} software \citep{2020ascl.soft09003H}. \textsc{oxkat} is a set of Python-based scripts that perform the traditional first-generation calibration (1GC), flagging, and second-generation calibration (2GC) using astronomy tools such as \textsc{casa} v5.6 \citep{2007ASPC..376..127M} and \textsc{wsclean} v3\footnote{\url{https://sourceforge.net/p/wsclean/wiki/Home/}} \citep{2014MNRAS.444..606O}. The target field did not have any problematic bright sources, hence we did not perform third-generation calibration (3GC), which includes correcting for direction-dependent effects.
\par
To compress the data and increase the processing speed, the data were averaged from 4096 channels to 1024 channels. We then proceeded to 1GC, where all the steps were performed using \textsc{casa}. We started with flagging the known RFI bands manually by specifying the frequency bands. We then applied automated flagging using \textsc{casa} tasks such as \textsc{rflag} and \textsc{tfcrop}. The primary and secondary calibrator models were used to calibrate the target field, and the calibrators and target were split into separate Measurements Sets (MS). 
\par
After applying the 1GC calibration solutions to the target field, the data were flagged using \textsc{tricolour} v0.1.7\footnote{\url{https://github.com/ska-sa/tricolour}} \citep{2022arXiv220609179H}. The final flagged percentage of the target data is 50.2$\%$. The target was imaged using \textsc{wsclean} with an auto-mask threshold, a Briggs robust weighting \citep{1995PhDT.......238B} of -0.3, and eight sub-band channels. The data was re-imaged in \textsc{wsclean} using the blind mask, with the final 1GC image then used to predict the sky model for the target field. This model was then used for self-calibration. The final step of the data reduction was 2GC, where one cycle of phase and amplitude self-calibration was performed using \textsc{casa} before a 2GC image was made. This final image's central frequency is 1.3\,GHz, and it has a rms of 8.5\,$\mu$Jy/beam near the cluster centre. The synthesised beam is  6.4$\arcsec$ $\times$ 5.9$\arcsec$, with a position angle of 61.7\,degrees. The full field of view is shown in Figure \ref{fig:fullfield}, with a zoom-in on the cluster region in Figure \ref{fig:bulletreg}.

\subsection{Compact source subtraction}
Prior to measuring the flux densities of the diffuse emission, we subtracted compact emission from all sources brighter than 25\,$\mu$Jy in a fixed region with a radius of 5$\arcmin$ centred at the cluster's SZ centre, which covered the extent of the diffuse emission. The compact source subtraction, which was applied in the uv-plane, was done in the to ensure that the flux density measurements for the diffuse emission had no contamination from the embedded compact emission. We also remove compact sources to ensure that they do not blend with the extended emission when we produce low resolution maps. The source subtraction was implemented using \textsc{crystalball} v0.3.1\footnote{\url{https://github.com/paoloserra/crystalball}} and \textsc{msutils} v1.2.0\footnote{\url{https://github.com/SpheMakh/msutils}}. \textsc{crystalball} uses a source list from \textsc{wsclean} to create a model column, in the MS file, that only contains the tagged compact sources. \textsc{msutils} then subtracts the newly created column from the column that contains the calibrated data and creates a new column containing the compact-source-subtracted visibilities. We then used the resulting visibilities to produce the low resolution map.

\begin{table}
    \centering
	\caption{Summary of the MeerKAT observations and the full resolution image properties.}
	\label{tab:observations}
	\begin{tabular}{lc} 
		\hline
		Central frequency (MHz) & 1283 \\
		$\Delta \nu$ (MHz) & 856\\
		No. of antennas & 61 \\
		Observing date(Y-M-D) & 2018-06-24 \\
		Amplitude calibrators & J0408$-$6545 $\&$ \\
		&J1331$+$3030 \\
		Phase calibrators & J0825$-$5010 $\&$ \\
		&PKS 0647$-$475 \\
		On-source time (hrs) & 7.7 \\
		$t_{\rm int}$ (sec) & 8 \\
		Flagged ($\%$) & 50.2 \\
		$\theta_{\rm synth}, \rm p.a.$ (\arcsec $\times$ \arcsec, \degree) & 6.4 $\times$ 5.9, 61.7\\
		$\sigma_{\rm rms}$ ($\mu$Jy/beam) & 8.5\\
		\hline
	\end{tabular}
\end{table}

\subsection{Low resolution maps}
\label{subsec:lowres}
\begin{figure}
	\includegraphics[width=\columnwidth]{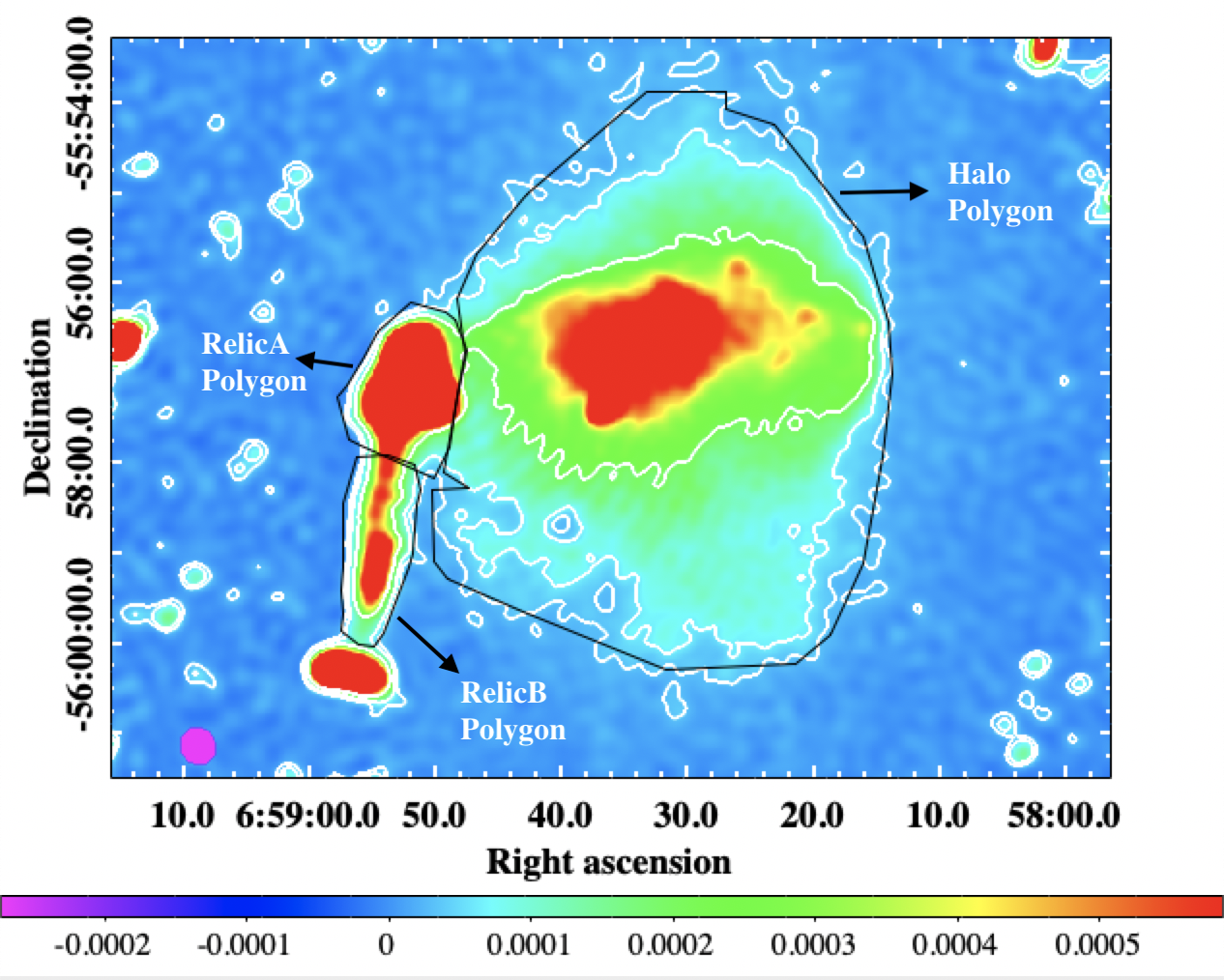}
    \caption{ MeerKAT's low resolution image of the diffuse emission. The rms of the image is 10.1 $\mu$Jy/beam. The beam, indicated by the magenta ellipse, has a size of 13.1$\arcsec$ $\times$ 11.7$\arcsec$ and the position angle of the beam is 120.6$\degree$. The black polygons indicate the regions that were used to calculate the flux density of the diffuse emission. The contour levels are [3,6,20]$\times \sigma$ where $\sigma$ = 10.1 $\mu$Jy/beam.}
    \label{fig:bulletreg2}
\end{figure}

We produce a low resolution map to enhance the surface brightness of the fainter regions of the large-scale diffuse emission. We imaged the target field using a uv-range $\leq$ 13\,k$\lambda$ which corresponds to a projected size of $\sim$119\,kpc at the cluster redshift, and a Briggs weighting robust parameter of 0. The diffuse emission in this cluster is extremely bright, and the flux density of the brightest regions is also captured in longer baselines. Hence, we used a higher $uv$-cut to ensure that we capture all the flux. We use the resulting image to measure flux densities of the diffuse radio sources. We use polygon regions guided by the 3$\sigma$ contours to create regions in which we measure the flux densities of the radio halo and relic. These regions are indicated by black polygons in Figure \ref{fig:bulletreg2}. We then use \textsc{ds9}'s \textsc{radioflux} v1.2\footnote{\url{https://github.com/mhardcastle/radioflux}} plug-in to measure the flux densities. The uncertainty on the measured flux densities is given by $\Delta S = \sqrt{(\delta S \times S)^2 + N \sigma^2} \;,$ where $S$ is the measured flux density, $\delta S$ is the flux calibration uncertainty which we assume to be 5$\%$ for MeerKAT \citep{2021arXiv211105673K}, $N$ is the number of beams within the region that the flux density was measured, and $\sigma$ is the local rms noise of the image.  

\subsection{Spectral Index Map}
\label{subsec:spec}
MeerKAT's sensitivity and wide bandwidth enables us to produce in-band spectral index maps from the single observation. To perform spectral index studies of the radio halo and relic in the Bullet cluster, we divided the compact source subtracted MS into two sub-bands with a bandwidth of $\sim$384\,MHz each. The central frequencies of the lower and upper sub-bands are 1.1\,GHz and 1.5\,GHz, respectively. We imaged each band using the multi-frequency mode in \textsc{wsclean} \citep{2017MNRAS.471..301O} and uv-matched both sub-bands to an inner and outer cut of 0.1<$uv$< 35\,k$\lambda$. The synthesised beam of the upper sub-band image was set to be the same as the synthesised beam of the lower sub-band image. Each image was primary beam corrected using \textsc{katbeam} v0.1\footnote{\url{https://github.com/ska-sa/katbeam}}. We use the sub-band images to calculate the in-band integrated spectral index. We calculate the corresponding uncertainties using the error propagation method. The spectral index maps were produced using Broadband Radio Astronomy Tools \citep[BRATS,][]{2013MNRAS.435.3353H,2015MNRAS.454.3403H} v2.6.3. To generate the spectral index maps, \textsc{brats}\footnote{\url{http://www.askanastronomer.co.uk/brats/}} uses the power law in the form of S$_{\nu}\, \propto \, \nu^{-\alpha}$ to produce the spectral index map values. This is performed using a pixel-by-pixel weighted least squares method, where the weights are the inverse variance. \textsc{brats} has an inbuilt flux density measurement function which requires the user to input the percentage calibration errors for the flux density uncertainty approximations, in our case we used 5$\%$. 
\par
To produce the spectral index map along with the uncertainty map, we load the sub-band maps into \textsc{brats}. We then load a \textsc{ds9} region file encompassing the extended diffuse emission along with an off-source background region (for details of the region definitions, see Section \ref{subsec:lowres}). \textsc{brats} uses the background region to calculate the rms noise of the low resolution maps. We set the source detection limit to be five times the rms noise. To produce spectral index maps, \textsc{brats} uses adaptive regions. These regions are set by a function written to group pixels into regions based on a specified set of parameters. We set the signal-to-noise parameter to be 6 and use the default settings for the rest of the parameters. In our analysis, we found that increasing the signal-to-noise parameter lowered the uncertainties associated with the spectral index map, resulting in a more robust map. 

\section{Results}
\label{sec:results} 
We detect both the known radio halo and relic and a new diffuse radio source in the MeerKAT $L$-band image of the Bullet cluster. We detect fainter larger scale diffuse emission compared to the ATCA observations \citep{2000ApJ...544..686L,2014MNRAS.440.2901S,2015MNRAS.449.1486S,2016Ap&SS.361..255M}. We discuss our findings below.

\begin{figure*}
	\includegraphics[scale=0.9]{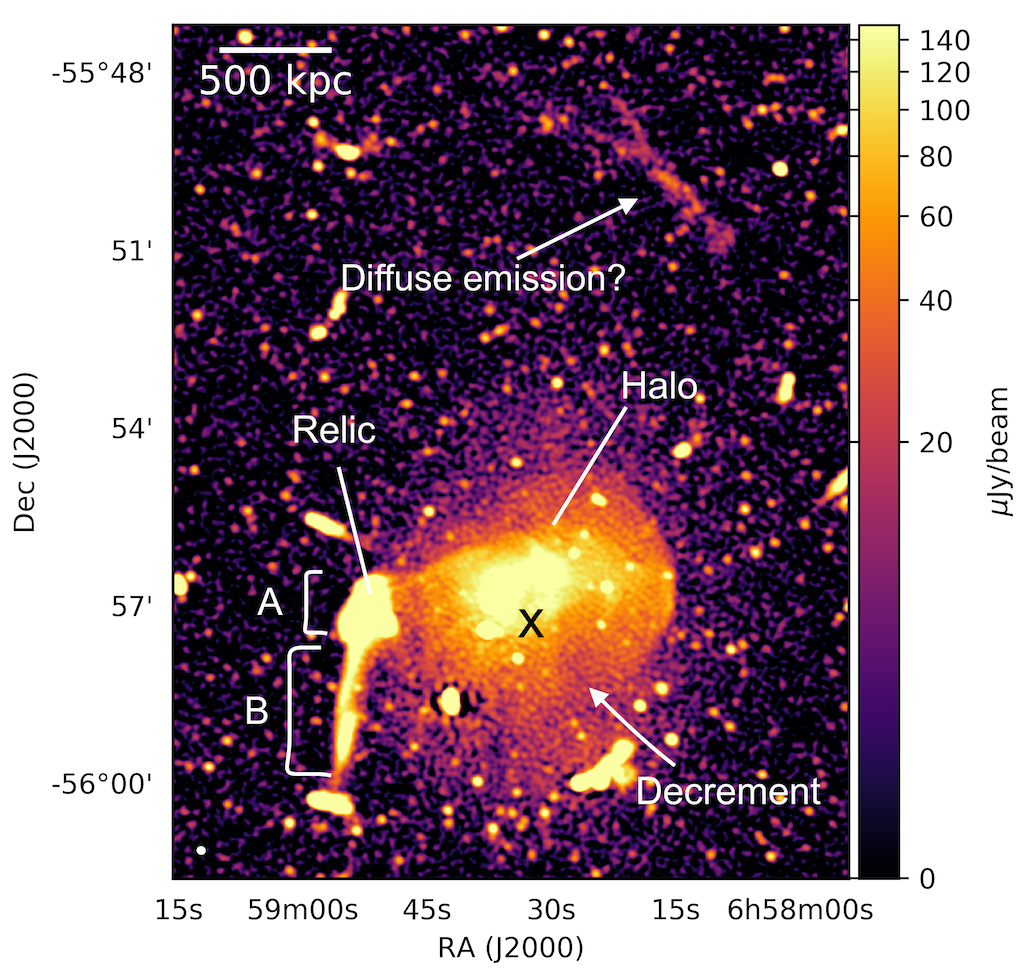}
    \caption{ Full resolution MeerKAT 1.3\,GHz image of the Bullet cluster. The rms of the image is $\sigma$ = 8.5\,$\mu$Jy/beam. The synthesized beam of the image is 6.4$\arcsec$ $\times$ 5.9$\arcsec$, 61.7$\degree p.a$, it is indicated by the filled white ellipse in the lower left corner of the image. We label the radio halo, relic, newly detected diffuse emission, and decrement region within the halo. The black cross indicates the SZ cluster centre.}
    \label{fig:bulletreg}
\end{figure*}

\subsection{Radio Halo}
The full-resolution MeerKAT $L$-band image of the Bullet cluster and labels of the diffuse radio sources is presented in Figure \ref{fig:bulletreg}. We extracted the diffuse emission properties from the low-resolution image, indicated by the contours in Figure \ref{fig:bulletgrid}. The radio halo emission is connected to the relic, hence we use the region where the relic's spectral index flattens ($\sim$0.6) as the halo boundary. The detected radio halo has an integrated flux density of 100.7\;$\pm$\;5.0\,mJy. The corresponding halo radio power is (34.4 \;$\pm$\; 1.1)$\times$10$^{24}$\,WHz$^{-1}$, which makes it one of the most powerful radio halos known to date. The largest angular size (LAS) of  the halo is 6.2$'$ $\times$ 4.7$'$ which corresponds to a largest linear size (LLS) of 1.6\,Mpc $\times$ 1.3\,Mpc.  MeerKAT's dense short baselines and sensitivity enables us to detect a larger north-south extent of the radio halo in comparison to previous observations \citep{2000ApJ...544..686L,2014MNRAS.440.2901S}. The observational properties of the structures are summarised in Table \ref{bulletradio-results}.
\par
Previous ATCA observations detect the brightest region of the radio halo, which overlaps with the 6$\sigma$ contours of the low resolution image (see Figure \ref{fig:bulletgrid}). In these observations, the radio halo is elongated in the east-west direction, which is along the merger axis. MeerKAT detects a fainter region of the radio halo, which results in the halo having a spherical morphology. However, the western region of the radio halo is still enclosed within the pronounced X-ray `bow-shock' envelope, as \citet{2014MNRAS.440.2901S} found in their paper for the ATCA observations. This shape indicates that X-ray shocks influence the shape of the diffuse radio emission. The extended faint halo emission appears to have a decrement below the bow shock region, as labelled in Figure \ref{fig:bulletreg}. This faint emission, including the decrement, traces the X-ray surface brightness (see Figure \ref{fig:bulletgrid}). The decrement pattern has been observed in clusters where the halo has substructures, such as in the Toothbrush cluster \citep{2016ApJ...818..204V,2020A&A...636A..30R}. This region is at the periphery of the protruding subcluster, and may be an indication of low electron density. However, comprehensive dynamical studies are required to determine if the fainter emission and decrement relate to any merger activity in the peripheral region of the cluster.

\subsection{Radio Relics}
As detected in \citet{2015MNRAS.449.1486S}, the morphology of the radio relic resembles a `toothbrush' which connects to the radio halo (see Figure \ref{fig:bulletreg}). MeerKAT's sensitivity once again allows us to trace a slightly extended length compared to the previous report. The radio relic has a LAS of  3.7$'$, corresponding to a  LLS of 988\,kpc. The integrated flux of the radio relic is 107.0\;$\pm$\;5.4\,mJy. The corresponding relic radio power is (43.5 \;$\pm$\; 0.5)$\times$10$^{24}$\,WHz$^{-1}$, which also makes the relic one of the most powerful relics known. The relic properties are summarised in Table \ref{bulletradio-results}. Similar to \citet{2015MNRAS.449.1486S}, we observe that region A of the radio relic is much brighter than region B. This varying brightness hints that region A might be connected to the remnants of a radio galaxy. This scenario supports the re-acceleration theory, which postulates that the cosmic ray electrons in radio relics come from a pre-existing pool of relativistic electrons.

\par
\begin{figure}
	\includegraphics[width=\columnwidth]{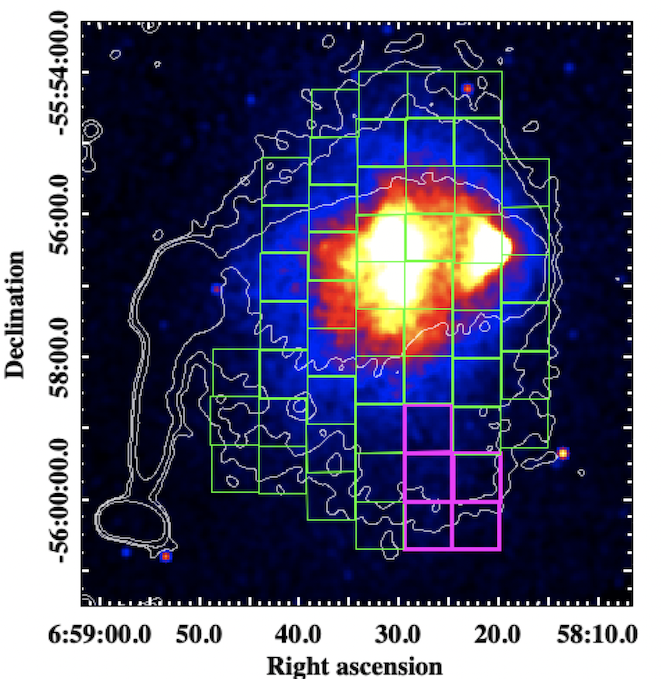}
    \caption{X-ray image of the Bullet cluster obtained from archival \textit{Chandra} ACIS-I observations by \citet{2006ESASP.604..723M}. The overlaid contours are from MeerKAT's low resolution image. The contour levels are [3,6,20]$\times \sigma$ where $\sigma$ = 10.1 $\mu$Jy/beam. The beam size of the low resolution image is 13.1$\arcsec$ $\times$ 11.7$\arcsec$ and the position angle of the beam is 120.6$\degree$. The grid shows the region which was used to perform the point-to-point analysis. All the cells were used for the MeerKAT analysis, while the magenta cells were excluded for the ATCA analysis.}
\label{fig:bulletgrid}
\end{figure}

We report a new detection of diffuse emission that is 2\,Mpc away from the SZ cluster centre, indicated in Figure \ref{fig:bulletgrid}. We searched for optical counterparts of the source in the Digitized Sky Survey-II \citep[DSS-II][]{2000ASPC..216..145M} image, since the region is not covered by archival Hubble Space Telescope observations. The detected source has no optical counterpart and is not connected to an individual radio galaxy. We measured the integrated flux density of the source to be 1.8\;$\pm$\;0.1\,mJy. It has an elongated morphology with a folk like structure on one end. It has a LAS of 3.4$\arcmin$, corresponding to a LLS of 0.9\,Mpc. The Bullet cluster is not paired with any neighbouring cluster, hence this reduces the chances of this source being associated with filaments. The detected source is $\approx$2\,Mpc away from the cluster's SZ centre. Hence, based on its properties, we propose that this could be a faint radio relic. Cosmological simulations \citep{2018ApJ...857...26H} predict 'equatorial shocks' that are generated by shock waves in the equatorial plane as a result of initial collision during a cluster merger. These shocks are formed much earlier than the shocks in the direction of the merger (axial shocks), hence the distance between equatorial shocks and the cluster centre is expected to be further than that for axial shocks. Equatorial shocks are rare and have been only observed at X-ray wavelength in one cluster \citep{2019NatAs...3..838G}. We propose that if this source is a relic, then it could be a candidate of a radio counterpart of an equatorial shock.

\begin{table*}  
    \begin{center}
    \begin{threeparttable}
    \caption{Summary of radio analysis of all the diffuse radio sources in the Bullet cluster. The columns are as follows. (1) Name of the diffuse radio source, (2) Integrated flux density, (3) Integrated in-band spectral index, (4) Radio power at 1.4 GHz, (5) Largest Angular Size, and (6) Largest Linear Size.}
    \begin{tabular}{lrrrrccl}
\hline
Source & Integrated Flux Density  & $\alpha^{1.1GHz}_{1.5GHz}$ & P$_{1.4GHz}$ & LAS & LLS &  \\
 & (mJy)   &  & (10$^{24}$ WHz$^{-1}$ ) & (arcmin)  & (kpc) & \\ \hline \hline
Halo & 100.7\;$\pm$\;5.0 & 1.1\;$\pm$\;0.2 & 34.4\;$\pm$\;1.1 &  6.2$\times$4.7 & 1632$\times$1250&\\
Relic region A & 98.6\;$\pm$\;4.9 & 1.1\;$\pm$\;0.1 & 40.0\;$\pm$\;1.3 & 1.6 & 432 &\\
Relic region B & 8.2\;$\pm$\;0.4 & 1.2\;$\pm$\;0.2 & 3.3\;$\pm$\;0.5 &  2.1 & 556 &\\
Relic & 107.0\;$\pm$\;5.4 & 1.1\;$\pm$\;0.2& 43.5\;$\pm$\;1.4 &  3.7 & 988&\\
candidate Relic& 1.8\;$\pm$\;0.1 & 1.6\;$\pm$\;0.9 & 0.6\;$\pm$\;0.01 &3.4 & 895\\
\hline    
\label{bulletradio-results}
\end{tabular}
\end{threeparttable}
\end{center}
\end{table*}

\subsection{Spectral index properties of the radio halo and relic}
\label{sec:specresults}
MeerKAT's wide bandwidth (856\,MHz) enabled us to perform in-band spectral index studies of the radio halo and the relic. We used sub-band images centred at 1.1\,GHz and 1.5\,GHz to calculate the integrated spectral indices of the radio halo, the full relic, and the newly detected source. We only divide the bandwidth to two sub-bands because this allows us to get sufficient signal-to-noise per sub-band for the fainter regions of the diffuse emission. The integrated spectral indices of the halo, relic, and diffuse source are 1.1\;$\pm$\;0.2, 1.1\;$\pm$\;0.2, and 1.6\;$\pm$\;0.9, respectively. The integrated spectral index of the radio halo is in agreement with measurements from \citet{2014MNRAS.440.2901S}. For region A and B of the relic, we measure the integrated spectral index values to be 1.06\;$\pm$\;0.05 and 1.21\;$\pm$\;0.17, respectively. The integrated spectral index value for region A is in agreement with the measurement in \citet{2015MNRAS.449.1486S} ($\alpha^{3.1}_{1.1}$=1.07\;$\pm$\;0.03), while region B has a shallower integrated spectral index than in \citet{2015MNRAS.449.1486S} ($\alpha^{3.1}_{1.1}$=1.66\;$\pm$\;0.14). This indicates that the integrated spectral index of region B, which is fainter than region A, steepens at higher frequencies. 
\par
\begin{figure*}
\centering
\includegraphics[width=0.495\textwidth]{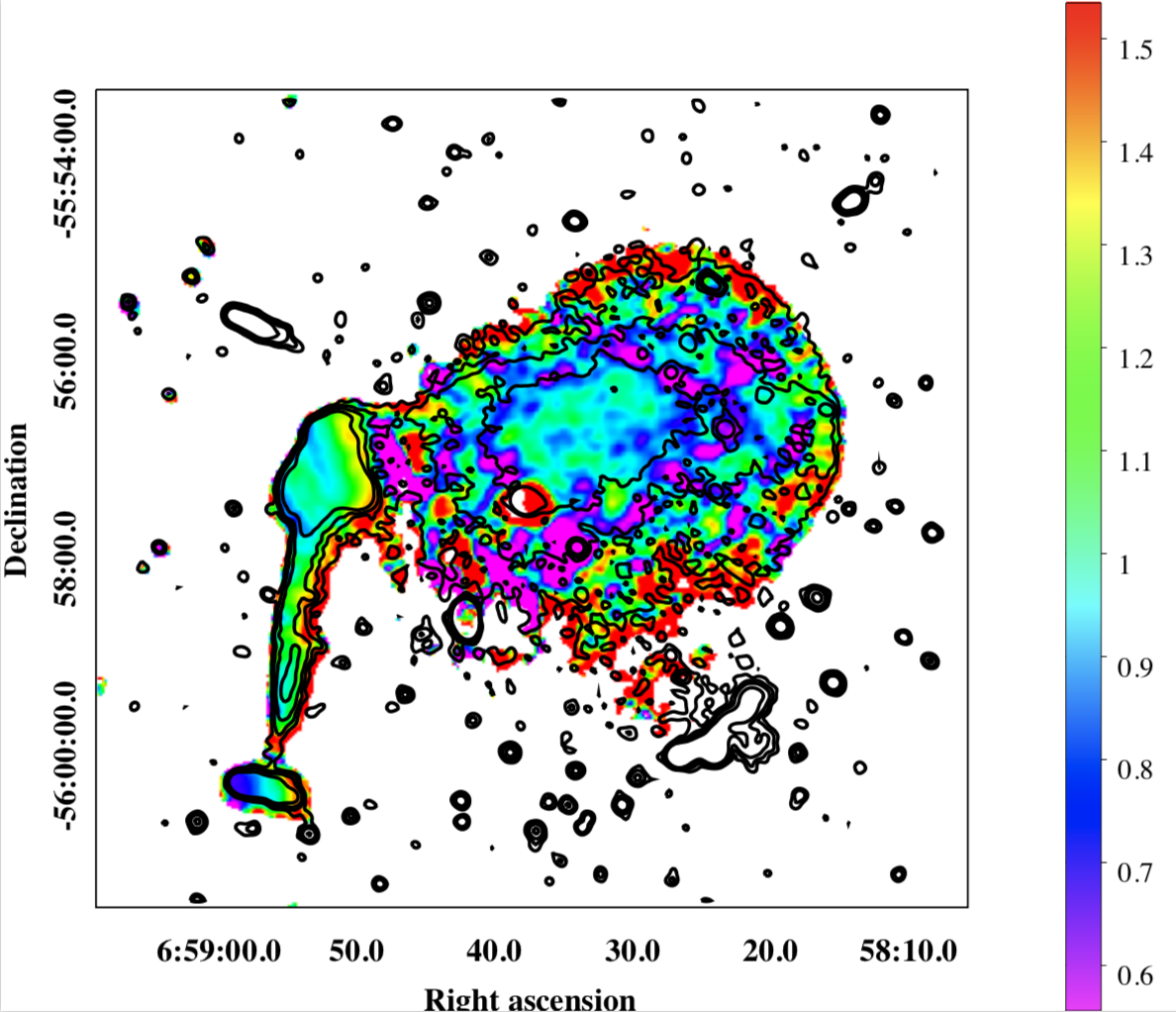}
\includegraphics[width=0.498\textwidth]{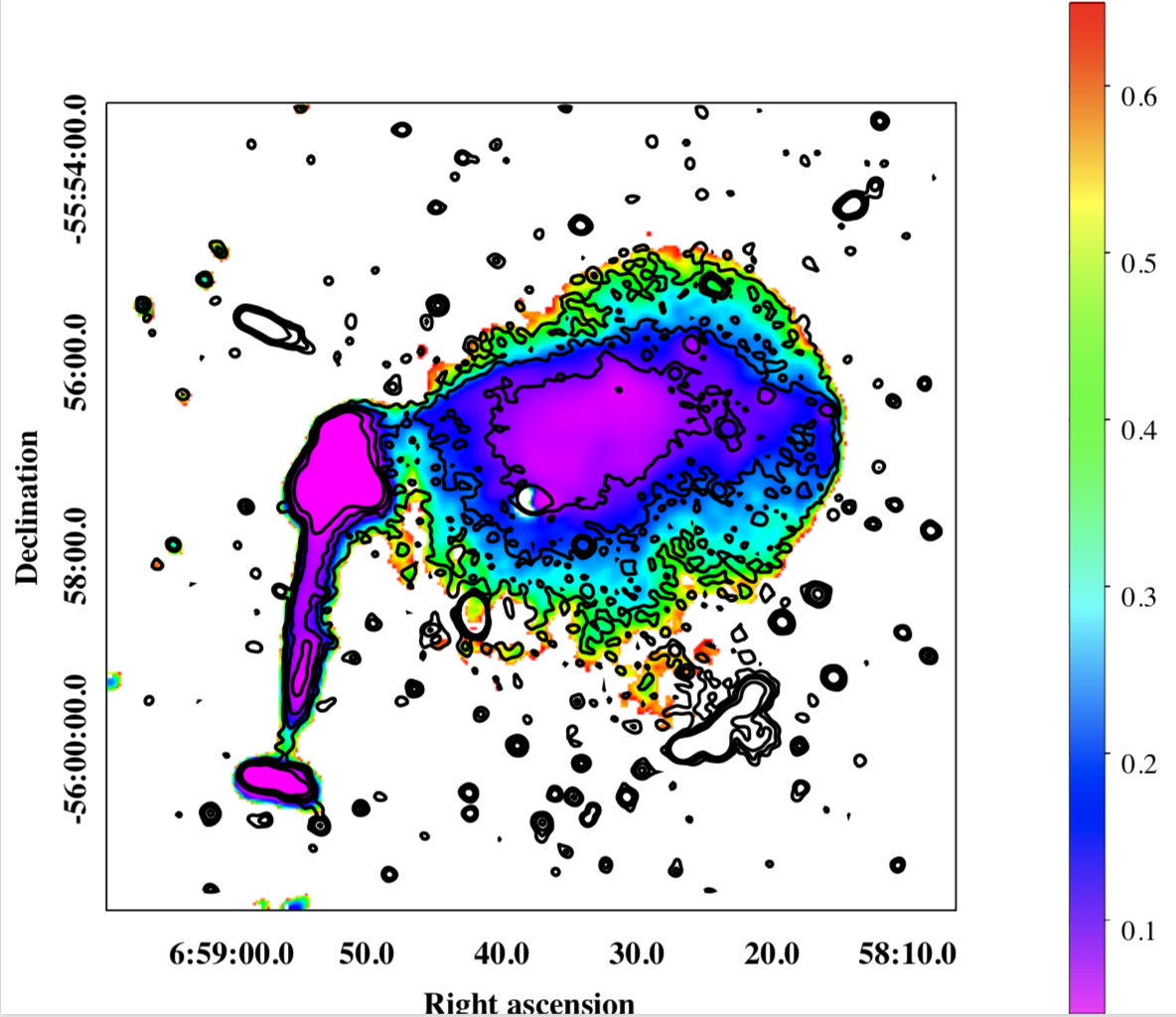}
\caption{\textit{Left:} In-band spectral index map of the radio halo and relic in the Bullet cluster. The map is produced from two sub-band full resolution images centred at 1.1\,GHz and 1.5\,GHz. The contours trace the diffuse emission from the full resolution (6.4$\arcsec$ $\times$ 5.9$\arcsec$) radio image, with levels of [3,6,12,]$\times \sigma$ where $\sigma$ = 8.5\,$\mu$Jy/beam. \textit{Right:} The corresponding spectral index uncertainty map.}
\label{fig:spec}
\end{figure*}

The spectral index map obtained from the  1.1\,GHz and 1.5\,GHz sub-band images is shown in Figure \ref{fig:spec}.  We were unable to produce a spectral index map of the newly detected emission region due to its faint surface brightness. The spectral index maps we obtain have a higher spatial resolution compared to those derived by \citet{2014MNRAS.440.2901S,2015MNRAS.449.1486S}. The spectral index map of the ratio halo also extends to the faint regions in the outskirts, which were previously not detected in ATCA observations. The brightest region of the halo, which in the central region, has a flatter spectrum with a minimal variation ($\alpha$ $ \approx$ 0.9 $-$ 1.0). We start to see more variation beyond the halo core ($\alpha$ $ \approx$ 0.6 $-$ 1.5), with the steepest part of the halo ($\alpha$ $\approx$1.5) being at the edges. The flatter spectrum core region indicates the presence of more energetic radiating particles, and/or of a larger value of the local magnetic field strength \citep{2004JKAS...37..315F,2020arXiv201214373R}. The radial steepening pattern has been previously reported for radio halos hosted by the Coma cluster, Abell\,665, and Abell\,2163 \citep{1993ApJ...406..399G,2004JKAS...37..315F}. In some instances, the steepening is observed due to unmatched uv coverage of the sub-band images. We ensured that we matched the sub-band uv coverage prior to creating the spectral index maps. Hence, it is unlikely that this pattern is due to data discrepancies. In the study of the Coma cluster \citet{2013A&A...558A..52B} ruled out the possibility that the SZ effect causes the observed steepening. The steepened spectrum coincides with the extended region where the radio halo emission is faint. This radial steepening pattern is predicted by primary models as a result of local magnetic field strength variation and/or varying relativistic electron's re-acceleration efficiencies.
\par
The spectral index map of the relic resembles that of a `toothbrush' relic \citep{2020A&A...642L..13R}. The relic's spectral index map indicates a spectral steepening in the E$-$W direction, towards the cluster centre. The spectral steepening feature is observed in most relics \citep{2006AJ....131.2900C,2015MNRAS.449.1486S,2020A&A...636A..30R}. This is caused by radiative losses in the downstream region.

\section{Discussion}
\label{sec:discuss} 
In Figure \ref{fig:bulletgrid}, it is evident that the larger halo region is tightly correlated with the overall shape of the X-ray surface brightness distribution. The low-resolution radio contours show that the radio halo's extension is larger in the north-south direction in comparison to the east-west direction, whereas previous studies by \citet{2000ApJ...544..686L} and \citet{2014MNRAS.440.2901S} reported that the radio halo was elongated along the E$-$W direction. The recovered halo's morphology traces the thermal X-ray emission. In our radio image, the extension along the E$-$W direction is 4.7\,arcminutes, while the north-south extension is 6.2\,arcminutes. Simulations of a 'bullet-like' clusters by \citep{2013MNRAS.429.3564D} showed radio halo emission elongated in the E$-$W direction even 0.22\,Gyrs after the first core passage. Although the core passage of the bullet occurred 0.1-0.2\,Gyrs ago \citep{2006ESASP.604..723M}, our detection of the extended faint emission shows that there is efficient re-acceleration of electrons even at outskirts of the cluster along the N$-$S direction.

\subsection{Radio halo vs X-ray brightness distribution }

The primary turbulent re-acceleration model predicts a connection between radio halos and thermal X-ray emission \citep{2014IJMPD..2330007B,2020A&A...640A..37I}. In this model, the merger induced turbulence is generated in the thermal background plasma and cosmic ray particles are re-accelerated and transported in that turbulence. This connection then gives rise to a spatial correlation between the radio and X-ray surface brightness. Correlation studies give insight to the interplay between non-thermal and thermal particles in the ICM, and further constrain existing radio halo formation theories. \citet{2001A&A...369..441G} were the first to investigate the relationship between  X-ray and  radio brightness. They used a subset of radio halos to quantify the linear correlation between radio and X-ray brightness in individual galaxy clusters. This linear relation is achieved under the assumption that the magnetic field strength and the energy density of the cosmic ray electrons are proportional to the thermal energy density.
\par
The radio and X-ray brightness relationship described by the power law, given by
\begin{equation}
\label{eq:bright}
\text{log}I_{\rm radio} = a + b\text{log}I_{\rm X-ray}.   
\end{equation}
The correlation probes the source of the cosmic ray particles \citep{2000A&A...362..151D,2001A&A...369..441G,2004MNRAS.350.1174B,2015ApJ...801..146Z}, where the slope of the best-fit line ($b$) indicates whether the magnetic field strength and cosmic ray energy density decline faster ($b>1$) or slower ($b<1$) than the thermal energy density. Turbulent re-acceleration models are able to produce both sub-linear and linear relations, while hadronic models only predict a super-linear correlation \citep{2014IJMPD..2330007B}. Previous studies have reported a sub-linear to linear relation for radio halos \citep{2001A&A...373..106F,2019A&A...628A..83C,2020ApJ...897...93B,2021A&A...646A.135R}. 
\par
\begin{figure*}
\centering
\includegraphics[scale=0.23]{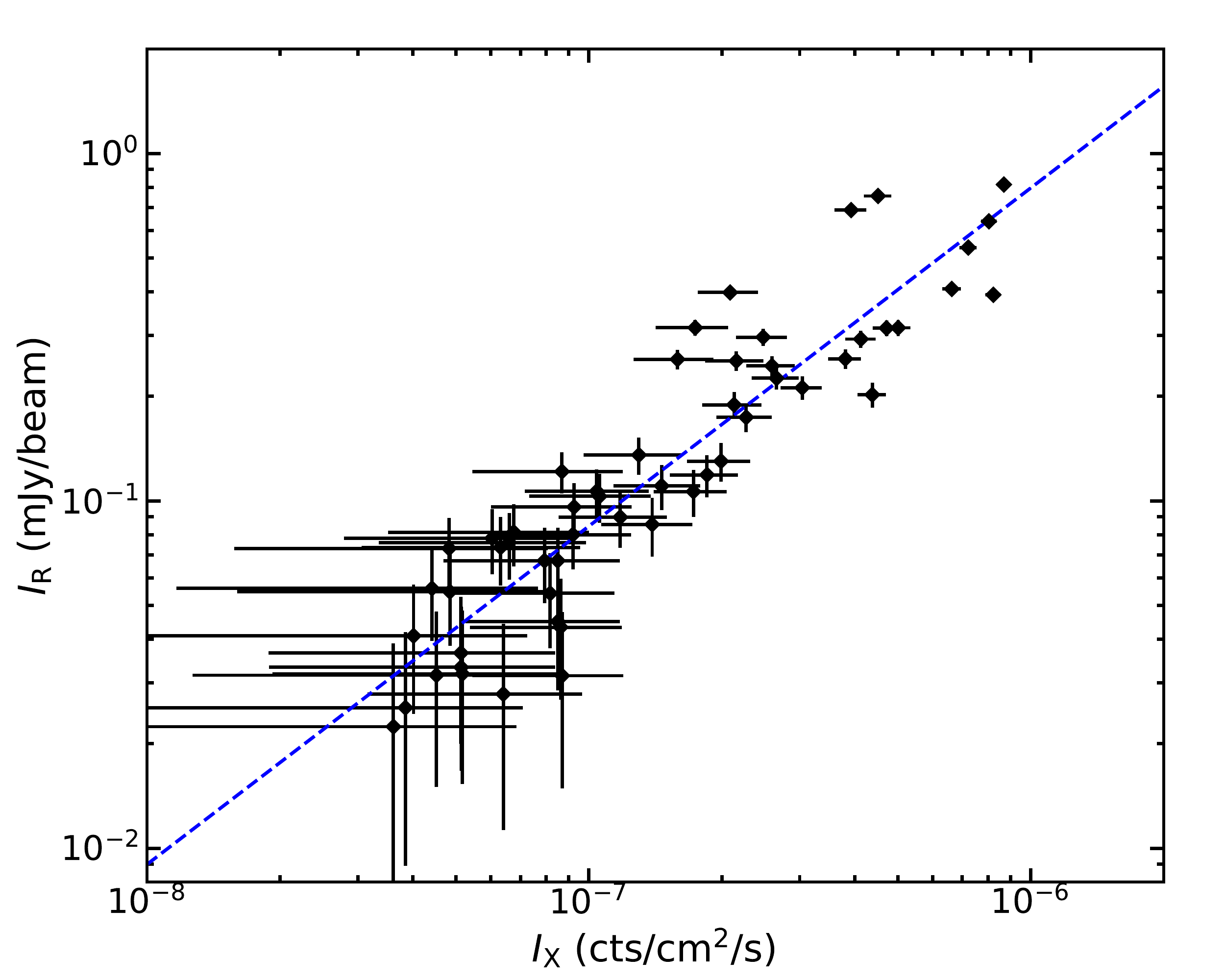}
\includegraphics[scale=0.23]{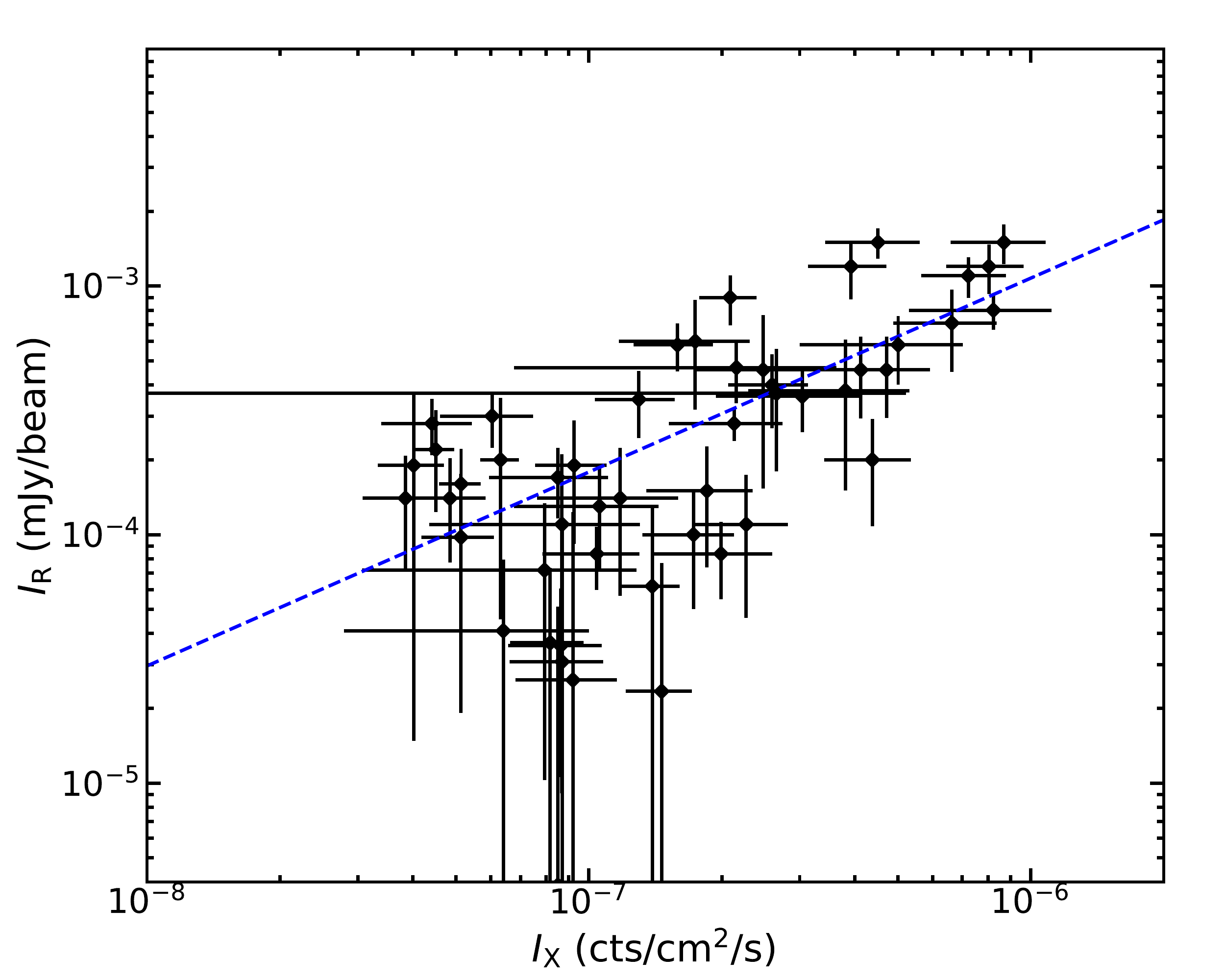}
\includegraphics[scale=0.23]{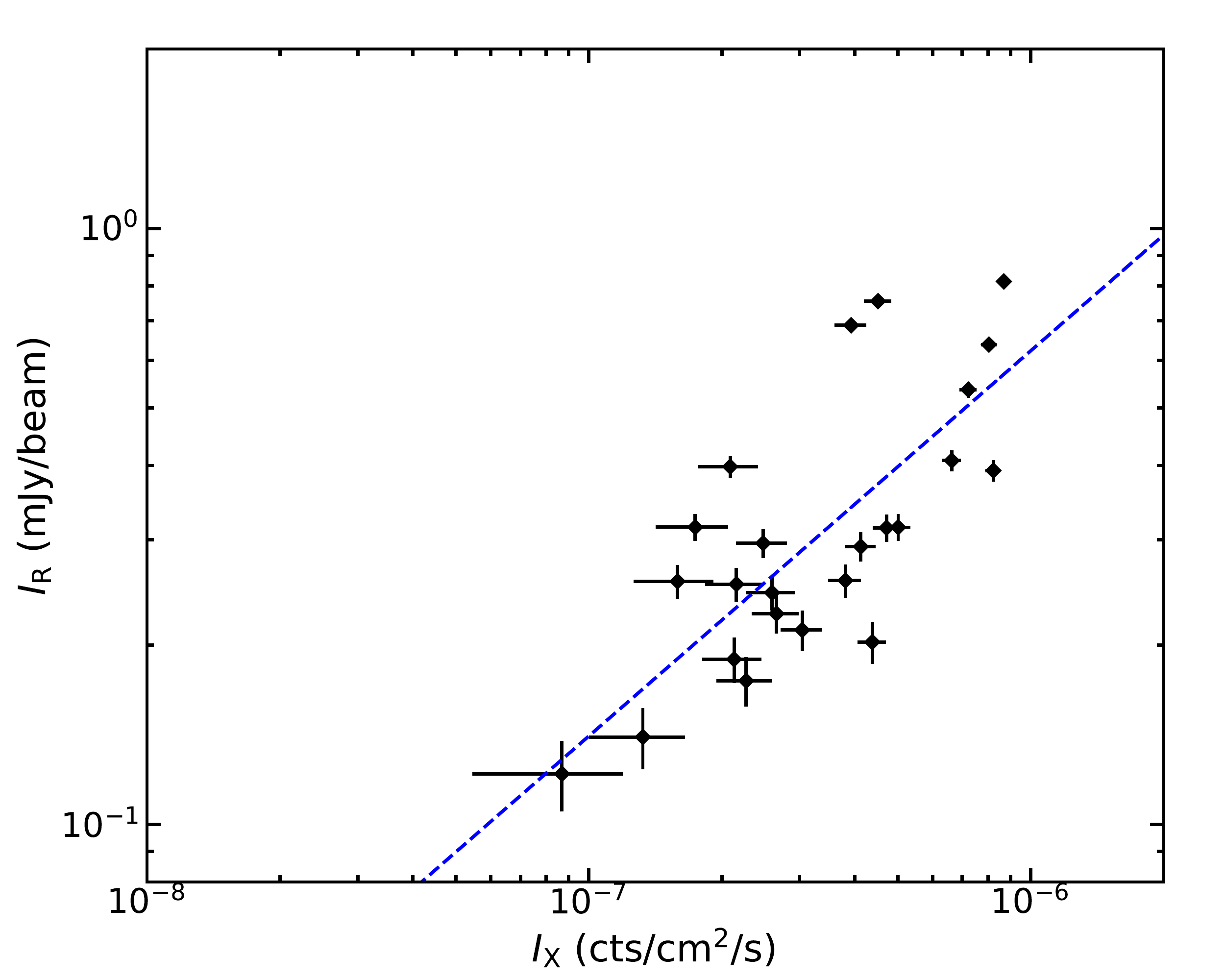}
\caption{Radio-X-ray correlation plots for MeerKAT and ATCA images. Each point indicates the correlation for individual cells, as produced by the \textsc{synage}++ package. The best-fit line is obtained using the MCMC method within \textsc{linmix}. To enhance the visibility of the error bars, the radio and X-ray error bars in the plots generated from the MeerKAT image were scaled by a factor of 5 and 10, respectively. \textit{Left:} The correlation plot produced using the low-resolution point-source-subtracted MeerKAT image with a central frequency of 1.3\,GHz. \textit{Middle:} The correlation plot produced using the ATCA 1.1$-$3.1\,GHz image \citep{2014MNRAS.440.2901S}. \textit{Right:} Radio-X-ray correlation plot using the brightest halo core in the MeerKAT image.}
\label{fig:fitscale}
\end{figure*}

To investigate the X-ray/radio correlation in the Bullet cluster, we use the archival X-ray image from the \textit{Chandra} ACIS-I observations (PI: Maxim Markevitch) and the MeerKAT low-resolution image. We first smoothed the X-ray image and re-gridded it to have the same number of pixels and resolution as the low-resolution radio image. We then created a gridded composite region of 55 cells with a dimension of 40$\arcsec\times$40$\arcsec$, 
indicated by green and magenta cells in Figure \ref{fig:bulletgrid}. The cell size was chosen to obtain a good compromise between high signal-to-noise ratio (within the cell) and good resolution. The size of the grid was guided by 3$\sigma$ contours of the low resolution radio image. We use the \textsc{synage++} package \citep{2011A&A...526A.148M} to calculate the point-to-point correlations in the region covered by the grid. The 'point' values are obtained from the individual cells. The uncertainties associated with each data point are calculated as the rms within the cells. The left panel of Figure \ref{fig:fitscale} shows the resulting correlation plot for the MeerKAT and \textit{Chandra} images. We use the \textsc{python} implementation of the \textsc{linmix} v0.1.0.dev1\footnote{\url{https://github.com/jmeyers314/linmix}} package to determine the best-fitting parameter values for the correlation \citep{2007ApJ...665.1489K}. \textsc{linmix} implements a Bayesian linear regression while accounting for intrinsic scatter, measurement uncertainties on both input variables, and derives an upper limit for the y-variable if it is not detected. Using the Markov Chain Monte Carlo \cite[MCMC][]{1970Bimka..57...97H} method performed within \textsc{linmix}, we take the mean of the posterior distribution as the best-fit slope. We use the \textsc{fitscalingrelation}\footnote{\url{https://github.com/mattyowl/fitScalingRelation}} [July 2022] package to plot the resulting correlation plots, each point represents the correlation per grid cell. The slope (\textit{b$_{1.3\rm GHz}$}) of the best-fit line is 0.97\;$\pm$\;0.06. We calculate the strength of the correlation using the Spearman and Pearson correlation coefficients ($r_s$, $r_p$). We obtain a Spearman and Pearson correlation coefficient of $r_s$=0.92 and $r_p$ = 0.84, respectively. We summarise our findings in Table \ref{tab:corr}.

\begin{table}
	\centering
	\caption{Radio halo and X-ray brightness correlation $b$ is the slope of the best-fit line, r$_s$ is the Spearman correlation coefficient, and r$_p$ is the Pearson correlation coefficient.}
	\label{tab:corr}
	\scalebox{0.92}{
	\begin{tabular}{lccccccl} 
		\hline
        Radio Image & Central Frequency & $b$ & r$_s$ & r$_p$ \\ \hline \hline
        MeerKAT full scale & 1.3 GHz &  0.97\;$\pm$\;0.06 & 0.92 & 0.84\\
        ATCA & 2.1 GHz & 0.78\;$\pm$\;0.12&0.66&0.79\\
        MeerKAT core & 1.3 GHz & 0.88\;$\pm$\;0.19 & 0.66&0.68\\
		\hline
	\end{tabular}
	}
\end{table}

\par
Our results indicate that there exists a linear correlation ($b\sim$1) between the radio and X-ray brightness. This outcome is contrary to that of \citet{2014MNRAS.440.2901S} who reported that they discovered no correlation and a low Spearman correlation strength of 0.4. To investigate this discrepancy, we use \textsc{linmix} to obtain best-fit values for the ATCA image (image obtained from Timothy Shimwell). We first used a grid the same size as in \citet{2014MNRAS.440.2901S} to reproduce their results, which indeed show no correlation. We then use the same grid we used for the MeerKAT correlation plot, excluding the magenta cells which overlap with the discrete sources that were not subtracted in the ATCA image. The resulting correlation plot for the ATCA and \textit{Chandra} images is shown in the middle panel of Figure \ref{fig:fitscale}. In this case we find a correlation of $b$ = 0.78\;$\pm$\;0.12, with a Spearman and Pearson correlation coefficient of $r_s$=0.66 and $r_p$ = 0.79, respectively. This indicates an increased correlation strength than previously reported. We probe whether the co-existence of different correlations is intrinsic or as a result of data quality. We use a smaller grid with 24 cells, covering the bright core, to re-calculate the best-fit value of the slope in the MeerKAT image. The correlation plot for the core region is shown in the right panel of Figure \ref{fig:fitscale}. The resulting slope has a best-fit value of 0.88\;$\pm$\;0.19, although flatter than the full-size radio halo fit, it is still consistent with the linear correlation and within the lower bound error bar. The Spearman and Pearson correlation coefficients for the core region are $r_s$=0.66 and $r_p$ = 0.68, respectively. Recent studies by \citet{2021MNRAS.508.3995B,2021A&A...654A..41R,2022arXiv220301958B} have reported disparities in correlation strengths of the halo core and the outskirts. Such a trend might be hinting that some regions of radio halos have a contribution from both hadronic re-acceleration mechanisms, as predicted by the hybrid models \citep{2014IJMPD..2330007B}. \citet{2022arXiv220301958B} cited that sub-linear regions have negligible contributions from the hadronic mechanisms. Studies of point-to-point analysis in larger samples of radio halos would be key to investigate if this trend exists in the majority of radio halos.


\subsection{Radio Mach Number}
The `bow' shock in the western region of the radio halo is well studied and constrained at X-ray wavelengths \citep{2002ApJ...567L..27M}. However, \citet{2015MNRAS.449.1486S} were the first to report a tentative shock in the region of the radio relic. They used archival \textit{Chandra} data to search for a surface brightness and temperature jump in the relic region. They discovered a clear surface brightness jump, but the temperature jump was not as distinct. They found the resulting Mach number to be $\mathcal{M}_X \approx$ 2.5$^{+1.3}_{-0.8}$.
\par
Radio relics are often associated with shock fronts and hence can be used as tracers of these shock regions. Considering the case where relics form through the DSA mechanism, the radio Mach number can be derived from the integrated spectral index with the assumption of stationary shock conditions in the ICM. The radio Mach number is then given by
\begin{equation}
\label{eq:mach}
\mathcal{M}_R = \left( \frac{2\alpha_{inj} + 3}{2\alpha_{inj} -1}, \right)^{1/2}
\end{equation}
where $\alpha_{inj}$ is the injection spectral index \citep{1962SvA.....6..317K,1987PhR...154....1B}. It can be related to the integrated spectral index in the following manner

\begin{equation}
\alpha_{int} = 0.5 + \alpha_{inj}.
\end{equation}
In Section \ref{sec:specresults} we calculated the integrated spectral index of the radio relic to be $\alpha_{int}$ =  1.1\;$\pm$\;0.2. From this value, we obtain a radio Mach number of $\mathcal{M}_R$= 4.6\;$\pm$\;0.9. The radio Mach number is higher than the estimated X-ray Mach number, as observed by multiple studies \citep{2016ApJ...818..204V,2018MNRAS.478.2218H,2019MNRAS.489.3905S}. X-ray Mach numbers are usually lower because they trace the average Mach number distribution while radio Mach numbers are sensitive to higher Mach numbers constrained in a small region of the shock's surface. \citet{2021MNRAS.506..396W} used simulations to investigate the discrepancies between the radio and X-ray Mach numbers. They concluded that radio Mach numbers are more robust as they are not affected by the orientation of the relic. Based on these findings, it would mean that the shock associated with the relic in the Bullet cluster is associated with the class of rare high Mach number ($\mathcal{M}_R$> 3) shocks.

\section{Conclusions}
In this paper, we studied the radio halo and relic in the Bullet cluster using MeerKAT $L$-band observations. After subtracting point sources, we were able to produce low resolution images that were significantly more sensitive than the previously reported observations. We detect a new diffuse radio source located north-west of the cluster centre, and $\approx$2\,Mpc away from the cluster's SZ centre. Due to the morphology of the source, we label it as a candidate relic. We also detect a decrement feature on the southern outskirts of the radio halo. The decrement feature and the radio halos `bow-shock' edge coincide with the X-ray surface brightness distribution. The sensitivity of MeerKAT enabled us to study the morphology of the halo and relic and produce their spectral index maps. The summary of our findings is as follows.
\par
(i) The integrated 1.3\,GHz flux density of the radio halo and relic was determined to be 100.7\;$\pm$\;5.0\,mJy and 107.0\;$\pm$\; 5.4\,mJy, respectively. We recovered a larger extent of both the radio halo and relic; hence, the recovered flux densities are significantly larger in comparison to the values quoted in \citet{2000ApJ...544..686L} and \citet{2014MNRAS.440.2901S,2015MNRAS.449.1486S}.
\par
(ii) We detected a fainter extension of the radio halo emission, which suggests lower relativistic electron density or a weaker magnetic field at the outskirts of the halo.
\par
(iii) The radio halo vs X-ray brightness correlation analysis shows a linear relation for the full extent of the halo. The slope of the best-fit line is 0.97\;$\pm$\;0.06 and a Spearman and Pearson correlation coefficient of $r_s$=0.92 and $r_p$ = 0.84, respectively.
\par
(iv) The relic's morphology resembles a `toothbrush' relic. Similar to \citet{2015MNRAS.449.1486S}, we found that the radio relic has a brighter `bulb' region, which is attached to a fainter tail. This indicates that the brighter region might have a remnant radio galaxy embedded. This radio galaxy could be the source of fossil electrons that seed the relic's formation.
\par
(v) Using sub-band images, we calculated the integrated spectral indices of the radio halo and relic to be 1.1 \;$\pm$\; 0.2 and 1.1 \;$\pm$\; 0.2, respectively. These spectral index values are in agreement with those reported by \citet{2014MNRAS.440.2901S,2015MNRAS.449.1486S}. Using the spectral indices, we calculated the radio power of the halo and the relic to be (34.4 \;$\pm$\; 1.1)$\times$10$^{24}$\,WHz$^{-1}$ and (43.5 \;$\pm$\; 0.5)$\times$10$^{24}$\,WHz$^{-1}$, respectively.  
\par
(vi) The spatial spectral index distribution of the radio halo indicates a radial steepening. The spectral index map of the relic indicates an E$-$W steepening pattern.
\par
(vii) There exists a linear correlation between the X-ray and radio brightness of the radio halo. However, the correlation strength decreases when considering only the bright inner core of the halo.
\par
(viii) The powerful relic is associated with a shock that has a high radio Mach number of $\mathcal{M}_R$= 4.6\;$\pm$\;0.9.
\par
Currently, L$-$band observations are the lowest frequency observations of the Bullet cluster. In future, UHF$-$band and lower frequency observations, could give insight on the full extent of the diffuse emission. This data would also allow us to produce a spectral index map of the newly detected radio source, which would confirm if it is indeed a radio relic associated with the cluster.

\section*{Acknowledgements}
The MeerKAT telescope is operated by the South African Radio Astronomy Observatory, which is a facility of the National Research Foundation, an agency of the Department of Science and Innovation. We acknowledge the assistance of the South African Radio Astronomy Observatory (SARAO) science commissioning and operations team (led by Sharmila Goedhart). We thank Maxim Markevitch for kindly sharing the X-ray surface brightness image. We also thank Timothy Shimwell for sharing the ATCA radio image and for the valuable input. We acknowledge the financial assistance of the SARAO towards this research. KM acknowledges support from the National Research Foundation of South Africa.



\section*{Data Availability}
The raw visibilities of the MeerKAT 1.3\,GHz observation of the Bullet cluster is accessible online from the MeerKAT Galaxy Clusters Legacy Surveys 
(MGCLS) data archive\footnote{\url{https://archive.sarao.ac.za/proposalid/SSV-20180624-FC-01/}} under the proposal ID SSV-20180624-FC-01. The data derived in this study are available upon reasonable request from the corresponding author.


\bibliographystyle{mnras}
\bibliography{Bullet_MeerKAT_MNRAS} 



\appendix

\section{Full Field of view and Low resolution images of the Bullet Cluster}
\begin{figure*}
    \centering
	\includegraphics[width=1.15\textwidth]{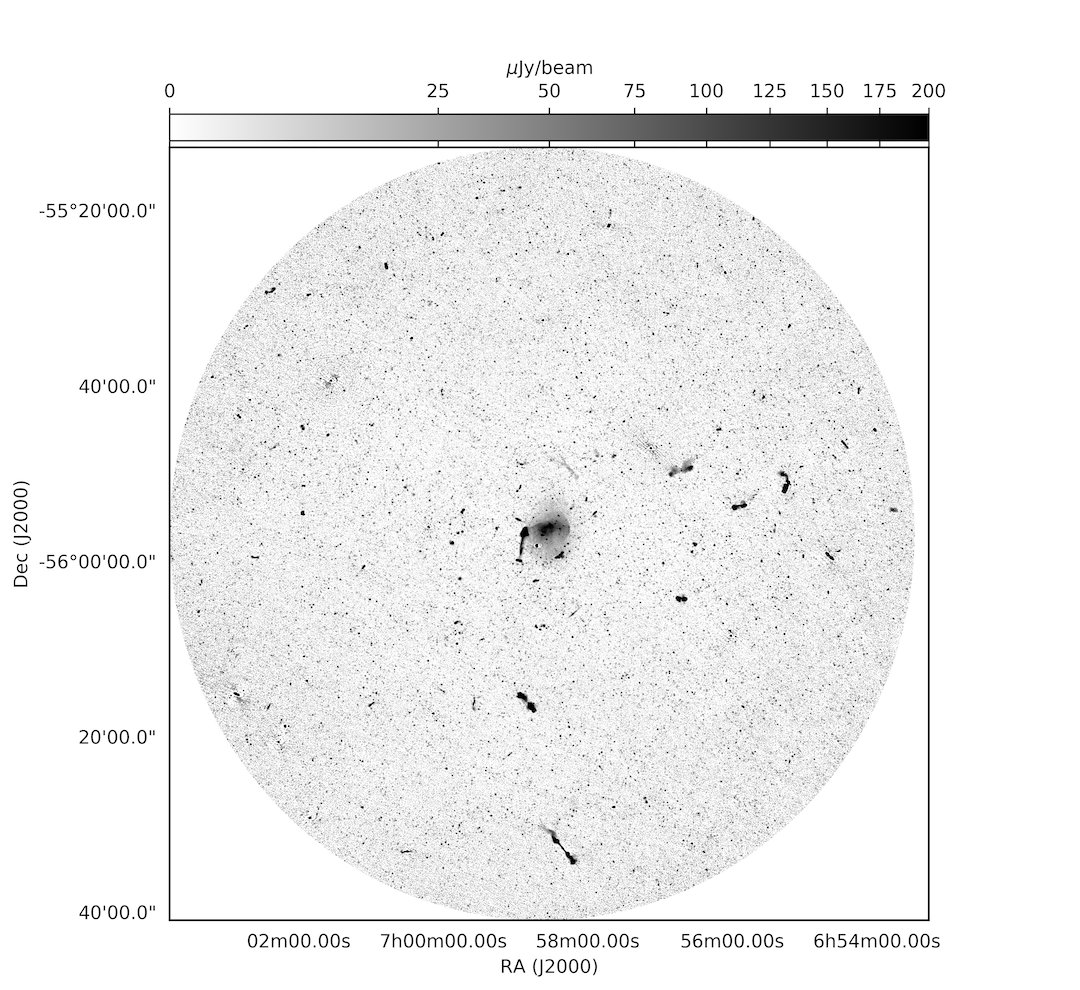}
	\caption{ The MeerKAT L-band primary beam corrected, full field of view image of the Bullet cluster. The rms noise of the image is 8.5 $\mu$Jy/beam.}
    \label{fig:fullfield}
\end{figure*}


\bsp	
\label{lastpage}
\end{document}